\documentclass{JHEP3}
\JHEPspecialurl{http://jhep.sissa.it/JOURNAL/JHEP3.tar.gz}
\usepackage{epsfig,multicol,bbm}
\newcommand\fverb{\setbox\pippobox=\hbox\bgroup\verb}
\newcommand\fverbdo{\egroup\medskip\noindent%
			\fbox{\unhbox\pippobox}\ }
\newcommand\fverbit{\egroup\item[\fbox{\unhbox\pippobox}]}
\newbox\pippobox

\newcommand{\be}{\begin{equation}}
\newcommand{\ee}{\end{equation}}
\newcommand{\bea}{\begin{eqnarray}}
\newcommand{\eea}{\end{eqnarray}}
\newcommand{\bc}{\begin{center}}
\newcommand{\ec}{\end{center}}
\newcommand{\bt}{\begin{tabular}}
\newcommand{\et}{\end{tabular}}
\newcommand{\bfig}{\begin{figure}}
\newcommand{\efig}{\end{figure}}
\newcommand{\bi}{\begin{itemize}}
\newcommand{\ei}{\end{itemize}}
\newcommand{\bleft}{\begin{flushleft}}
\newcommand{\eleft}{\end{flushleft}}
\newcommand{\bright}{\begin{flushright}}
\newcommand{\eright}{\end{flushright}}
\newcommand{\bpage}{\begin{minipage}}
\newcommand{\epage}{\end{minipage}}

\newcommand{\pip}{\ensuremath{\pi^+\,}}
\newcommand{\pim}{\ensuremath{\pi^-\,}}
\newcommand{\piz}{\ensuremath{\pi^0\,}}
\newcommand{\Ks}{\ensuremath{K_S \,}}
\newcommand{\Kl}{\ensuremath{K_L\,}}

\newcommand{\etap}{\ensuremath{\eta'\,}}

\renewcommand{\to}{\ensuremath{\rightarrow}}

\newcommand{\pbinv}{\ensuremath{\,{\rm pb}^{-1}}}

\newcommand{\MeV}{\ensuremath{\,{\rm MeV}}}
\newcommand{\fikskl}{\ensuremath{\phi\rightarrow\Ks\Kl}}
\newcommand{\fipippimpiz}{\ensuremath{\phi\rightarrow\pi^+\pi^-\pi^0}}
\newcommand{\fietag}{\ensuremath{\phi\rightarrow\eta\gamma\;}}

\newcommand{\etappp}{\ensuremath{\eta\rightarrow\pip\pim\piz}}
\newcommand{\etapizpizpiz}{\ensuremath{\eta\rightarrow\piz\piz\piz}}

\newcommand{\etatrepi}{\ensuremath{\eta\rightarrow3\pi}}

\title{\mathversion{bold}
Determination of $\eta\to\pi^+\pi^-\pi^0 $ Dalitz plot slopes and asymmetries with the KLOE detector. }
\author{The KLOE collaboration:\\
F.~Ambrosino,$^{e,f,*}$
A.~Antonelli,$^a$
M.~Antonelli,$^a$
F.~Archilli,$^a$
C.~Bacci,$^{j,k}$
P.~Beltrame,$^b$
G.~Bencivenni,$^a$
S.~Bertolucci,$^a$
C.~Bini,$^{h,i}$
C.~Bloise,$^a$
S.~Bocchetta,$^{j,k}$
F.~Bossi,$^a$
P.~Branchini,$^k$
R.~Caloi,$^{h,i}$
P.~Campana,$^a$
G.~Capon,$^{a}$
T.~Capussela,$^{a,}$\footnote{
\rm{Corresponding authors.}\\ \it{e-mail addresses}\rm{: fabio.ambrosino@na.infn.it (F. Ambrosino), tiziana.capussela@lnf.infn.it (T. Capussela), francesco.perfetto@na.infn.it (F. Perfetto)}},
~~F.~Ceradini,$^{j,k}$
F.~Cesario,$^{j,k}$
S.~Chi,$^a$
G.~Chiefari,$^{e,f}$
P.~Ciambrone,$^a$
F.~Crucianelli,$^h$
E.~De~Lucia,$^a$
A.~De~Santis,$^{h,i}$
P.~De~Simone,$^a$
G.~De~Zorzi,$^{h,i}$
A.~Denig,$^b$
A.~Di~Domenico,$^{h,i}$
C.~Di~Donato,$^f$
B.~Di~Micco,$^{j,k}$
A.~Doria,$^f$
M.~Dreucci,$^a$
G.~Felici,$^a$
A.~Ferrari,$^a$
M.~L.~Ferrer,$^a$
S.~Fiore,$^{h,i}$
C.~Forti,$^a$
P.~Franzini,$^{h,i}$
C.~Gatti,$^a$
P.~Gauzzi,$^{h,i}$
S.~Giovannella,$^a$
E.~Gorini,$^{c,d}$
E.~Graziani,$^k$
W.~Kluge,$^b$
V.~Kulikov,$^n$
F.~Lacava,$^{h,i}$
G.~Lanfranchi,$^a$
J.~Lee-Franzini,$^{{a,l}}$
D.~Leone,$^b$
M.~Martini,$^{a,e}$
P.~Massarotti,$^{e,f}$
~~W.~Mei,$^{a}$
S.~Meola,$^{e,f}$
S.~Miscetti,$^a$
M.~Moulson,$^a$
S.~M\"uller,$^a$
F.~Murtas,$^a$
M.~Napolitano,$^{e,f}$
F.~Nguyen,$^{j,k}$
M.~Palutan,$^a$
E.~Pasqualucci,$^i$
A.~Passeri,$^k$
V.~Patera,$^{a,g}$
F.~Perfetto,$^{e,f,*}$
M.~Primavera,$^d$
P.~Santangelo,$^a$
G.~Saracino,$^{e,f}$
B.~Sciascia,$^a$
A.~Sciubba,$^{a,g}$
A.~Sibidanov,$^{a}$
T.~Spadaro,$^{a}$
M.~Testa,$^{h,i}$
L.~Tortora,$^k$
P.~Valente,$^i$
G.~Venanzoni,$^a$
R.Versaci,$^a$
G.~Xu,$^{a,m}$.\\
\llap{$^a$}Laboratori Nazionali di Frascati dell'INFN, Frascati, Italy\\
\llap{$^b$}Institut f\"ur Experimentelle Kernphysik, Universit\"at Karlsruhe, Germany\\
\llap{$^c$}Dipartimento di Fisica dell'Universit\`a, Lecce, Italy\\
\llap{$^d$}INFN Sezione di Lecce, Lecce, Italy\\
\llap{$^e$}Dipartimento di Scienze Fisiche dell'Universit\`a  ``Federico II'', Italy\\
\llap{$^f$}INFN Sezione di Napoli, Napoli, Italy\\
\llap{$^g$}Dipartimento di Energetica dell'Universit\`a ``La Sapienza'', Roma, Italy\\
\llap{$^h$}Dipartimento di Fisica dell'Universit\`a ``La Sapienza'', Roma, Italy\\
\llap{$^i$}INFN Sezione di Roma, Roma, Italy\\
\llap{$^j$}Dipartimento di Fisica dell'Universit\`a ``Roma Tre'', Roma, Italy\\
\llap{$^k$}INFN Sezione di Roma Tre, Roma, Italy\\
\llap{$^l$}Physics Department, State University of New York at Stony Brook, USA\\
\llap{$^m$}Institute of High Energy Physics of Academia Sinica,  Beijing, China\\
\llap{$^n$}Institute for Theoretical and Experimental Physics, Moscow, Russia\\
}
\preprint{pre-print xxx}
\received{\today} 		
\revised{\today}
\accepted{\today}		
\abstract{
We have studied, with the KLOE detector at the DA$\Phi$NE $\Phi$-Factory, the dynamics of
the decay \etappp\ using $\eta$ mesons from the decay $\phi\to\eta\gamma $  for an integrated luminosity ${\mathcal L}$ = 450 \pbinv.
From a fit to the Dalitz plot density distribution we obtain a precise measurement of the slope parameters.
An alternative parametrization relates  the $\pip \pim \piz$ slopes to that for $\eta\to3\piz$ showing the consistency of KLOE results for both channels.
We also obtain the best confirmation of the $C$-invariance in the
\etappp\ decay.}
\def\ifm#1{\relax\ifmmode#1\else$#1$\fi}

\def\gam{\ifm{\gamma}} \def\to{\ifm{\rightarrow}}
\def\pip{\ifm{\pi^+}} \def\pim{\ifm{\pi^-}}
  \def\x{\ifm{\times}}

    \def\K{\ifm{K}} 
\def\Kb{\ifm{\rlap{\kern.3em\raise1.9ex\hbox to.6em{\hrulefill}} K}}
\def\ab{\ifm{\sim}}  \def\x{\ifm{\times}}  
\def\amp#1,#2,{\ifm{\langle#1|#2\rangle}}

\def\deg{\ifm{^\circ}}

   \def\plm{\ifm{\pm}}

   \def\etap{\ifm{\eta'}}    
\def\C{\ifm{C}}    
\def\figb#1;#2;{\parbox{#2cm}{\epsfig{file=#1.eps,width=#2cm}}}   \let\cl=\centerline
\def\figbc#1;#2;{\cl{\figb #1;#2;}}
\font\euler=eufm10 at 12pt    \def\Ma{\hbox{\euler M}}
\catcode`@=11 
\newdimen\z@ \z@=0pt 
\newskip\z@skip \z@skip=0pt plus0pt minus0pt
\def\m@th{\mathsurround=\z@}
\def\ialign{\everycr{}\tabskip\z@skip\halign} 
\def\eqalign#1{\null\,\vcenter{\openup\jot\m@th
  \ialign{\strut\hfil$\displaystyle{##}$&$\displaystyle{{}##}$\hfil
      \crcr#1\crcr}}\,}
\catcode`@=12 
\begin{document}
\section{Introduction}
The decay $\eta\to3\pi$ violates iso-spin invariance.
Electromagnetic contributions to the process are very small
\cite{Sutherland} and the
decay is induced dominantly by the strong interaction via the $u,d$ mass
difference.
The $\eta\to3\pi$ decay is therefore an ideal laboratory for testing chiral
perturbation theory, ChPT.
A three body decay\footnote{Both $\eta$ and $\pi$ are spinless, therefore
  there is no preferred direction.}
is fully described by two variables. We can choose two of the pion energies
($E_+,E_-,E_0$)
in the $\eta$ rest frame, two of the three two pion masses squared
($m^2_{+-},\ m^2_{-0},\ m^2_{0+}$)
also called ($s,t,u$). Note that $E_+$ is linear in $m^2_{-0}$ and so on,
cyclically.
We use the Dalitz variables, $X,Y$ which are linear combinations of the pion energies:
\begin{equation}\eqalign{X&= \sqrt3\,{E_+-E_-\over Q} = {\sqrt3\over 2m_\eta Q}\,(u-t)\cr
 Y& = 3\,{E_0-m_0\over Q}-1 = {3\over2m_\eta Q}\left((m_\eta-m_{\piz})^2-s\right)-1\cr}
\label{defXY}
\end{equation}
where $Q$ is the decay ``$Q$-value''.
The decay amplitude is given in \cite{BijGa02} as:
\begin{equation}
A(s,t,u)=\frac{1}{\Delta^2}\frac{m_K^2}{m_\pi^2}
\left(m_\pi^2-m_K^2\right)\frac{\Ma(s,t,u)}{3 \sqrt{3} F^{2}_{\pi}}
\label{ampli}
\end{equation}
where $\Delta^2\equiv(m_s^2-\widehat m^2)/(m_d^2-m_u^2)$ and
$\widehat m=(m_u+m_d)/2$ is the average $u, d$ quark mass. $F_\pi = 92.4$
MeV is the pion decay constant
 and $\Ma(s,t,u)$ must come from theory. From eq. \ref{ampli} it follows that the decay rate for \etappp\ is proportional to $\Delta^{-4}$.
The transition \etatrepi\ is therefore very sensitive to $\Delta$ if the amplitude \Ma\ were known.
At lowest order in ChPT:
\begin{equation}
\Ma(s,t,u) =\frac{3 s-4 m_{\pi}^{2}}{m_{\eta}^{2}-m_{\pi}^{2}}.
\label{ma}
\end{equation}
From eq. \ref{ma}, \cite{BijGa02} one finds $\Gamma^{\rm lo}\left(\etappp\right)=66$ eV to be compared with the measured width of  $295 \plm 16$ eV \cite{PDG}.
\noindent
A one-loop calculation within conventional chiral perturbation theory
(ChPT) \cite{GasLut85}, improves  considerably the prediction:
\begin{equation}
\Gamma^{\rm nlo}\left(\etappp\right)\simeq 167 \pm 50
\;\textrm{eV}.
\label{eq:theo2}
\end{equation}
\noindent
but is still far from the experimental value. Higher order corrections \cite{BijGho07}
help but do not yet bring agreement with measurements of both total rate and Dalitz plot slopes. Good agreement is found combining ChPT with a non perturbative coupled channels
 approach using the Bethe Salpeter equation \cite{Borasoy}.

Therefore  a precision study of the \etatrepi\ Dalitz plot, DP, is highly desirable. The amplitude squared
is expanded around $X=Y=0$  in power of $X$ and $Y$
\begin{equation}
|A(X,Y)|^2\propto1 + aY + bY^{2} + c X + dX^{2} + eXY +....
\label{eq:amp_tra}
\end{equation}
The parameters ($a,b,c,d,e,...$ ) can be obtained from a fit to the observed DP density and should be computed by the theory.
Any odd power of $X$ in $A(X,Y)$ implies violation of charge conjugation.

\section{The KLOE detector}
Data were collected with the KLOE detector at DA\char8NE \cite{DAFNE},
the Frascati $e^{+} e^{-}$ collider, which operates at a
center of mass energy $W =m_{\phi}\sim 1020$
MeV . The electron and  positron beams collide with a
crossing angle of $\pi - 25$ mrad, resulting in a small momentum
($p_{\phi}\sim$ 13 MeV/c in the horizontal plane) of the produced $\phi$
mesons.
The KLOE detector consists
of a large cylindrical drift chamber (DC), surrounded by a fine
sampling lead-scintillating fibers electromagnetic calorimeter (EMC)
inserted in a 0.52 T magnetic field.

The DC \cite{DC}, 4 m diameter and 3.3 m long, has full stereo
geometry and operates with a gas mixture of 90\% helium and 10\%
isobutane.
Momentum resolution is $\sigma(p_\bot)/p_\bot\leq
0.4\%$. Position resolution in $r - \phi$ is 150 $\mu$m and $\sigma_{z}\sim$ 2
mm. Charged tracks vertices are reconstructed with an accuracy of $\sim$ 3 mm.

The EMC \cite{EMC} is divided into a barrel and two endcaps, for a total
of 88 modules, and covers 98\% of the solid angle.
Arrival times of particles and space positions  of
the energy deposits are obtained from the signals collected at the two
ends of the calorimeter modules, with a granularity of $\sim$(4.4 x
4.4) cm$^{2}$, for a total of 2240 cells arranged in five
layers. Cells close in time and space are grouped into a calorimeter
cluster. The cluster energy $E$ is the sum of the cell energies, while
the cluster time $t$ and its position {\bf r} are energy weighted
averages. The respective resolutions are $
\sigma_E/E = 5.7\% / \sqrt{E\ ({\rm GeV})}$
and $\sigma_t = 57 \;\textrm{ps} / \sqrt{E\ ({\rm GeV})} \oplus 100\;
\textrm{ps}$.

The KLOE trigger \cite{TRIGGER} is based on the coincidence of at
least two  energy deposits in the EMC above a threshold that
ranges between 50 and 150 MeV. In order to reduce the trigger rate due
to cosmic rays crossing the detector, events with a large energy
release in the outermost calorimeter planes are vetoed.

\section{Signal selection and efficiency}
\label{Efficiencies}

This analysis refers to $\sim$ 450 pb$^{-1}$ collected at  DA\char8NE in
years 2001/02 corresponding to $\sim 1.4~ 10^9$  $\phi$ mesons produced.

At KLOE  $\eta$ mesons are produced through the radiative decay $\fietag$.
Accounting for the product of BR's:
${\rm BR}(\fietag)\times\!{\rm BR}(\etappp) \simeq 2.9 \x10^{-3}$
we expect about four millions of $\etappp$ events. A
larger Monte Carlo (MC) sample corresponding to about $5$ times the
amount of data  has been used to study efficiencies
and backgrounds.


Note that the recoil photon is almost monochromatic, with
$E_{\gam\ {\rm rec}}\sim 363$  MeV, well separated from the
softer  photons from $\piz$ decay.

A photon is defined as  an EMC cluster  not associated to a DC track.
We further require that $|(t-r/c)|<5\sigma_t$, where $t$ is the arrival time at the EMC, $r$ is the
distance of the cluster from interaction point, IP, $c$ is speed of light.
The events selection is performed through the following steps:
\begin{enumerate}
\item Events are first selected by a very loose offline  filter to remove machine
  background (FILFO) and an event  selection procedure (EVCL) assigning
 events into categories \cite{EVCL}.
\item We then require two opposite curvature tracks intersecting at a point (vertex) inside a cylinder with
 $r<4$ cm, $|z|<8$ cm centered at the IP. We require also three photons with $ 21\deg<\theta_{\gamma}<159\deg$ and
 $E_{\gamma}>10$ MeV. The angle between any photon pair must be $>18^{\circ}$ to remove split showers.
\item $\sum E_{\gamma} <800\MeV.$
\item A constrained kinematic fit is performed imposing 4-momentum
  conservation and $t=r/c$ for each photon. We retain events with a
  probability $P(\chi^{2})>1$\%, corresponding to $\chi^{2}<18$.
The fit significantly improves the photon energy resolution.
The  $\chi^2$ distribution is in reasonable agreement with MC
prediction, as shown in fig.\ref{chisq_eta_2}; varying the cut on
$P(\chi^{2})$ in the range [ 0.01, 0.15] $(\chi^2<18 to \chi^2<10)$ has no significant effect on the analysis results, see section \ref{Systematics}.
\FIGURE{
\cl{\figb chi2;6.5;\kern.5cm\figb chi2log;6.5;}
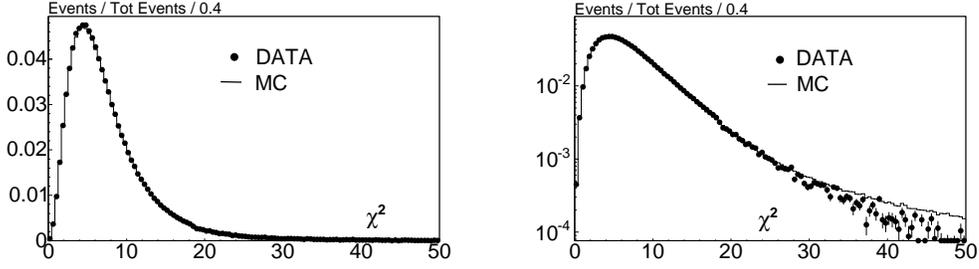
\caption{ $\chi^2$ distribution for the kinematic fit. Left: linear scale. Right: log scale.}
\label{chisq_eta_2}}
\item Finally we require:
\begin{enumerate}
\item  320 MeV$ < \!E_{\gamma,\ {\rm rec}}\! <$ 400 MeV for the recoil photon, to reduce residual background from \fikskl\ events.
\item $E_{\pip}+E_{\pim}< $550 MeV,  to reduce residual background from \fipippimpiz\ events.
\item $m$(\gam\gam) for the two softest photons must satisfy 110$<m_{\gamma\gamma}<$160
  MeV,  to reduce residual background from \etappp\ decays with $\piz \rightarrow e^+e^-\gamma$ and from $\phi\rightarrow\omega\piz$ with $\omega\rightarrow\pip\pim\piz$; and to eliminate the residual background from \fietag\ events with $\eta \rightarrow \pip \pim \gamma$.
\end{enumerate}
\end{enumerate}
The selection efficiency is determined with the MC
program \cite{EVCL} and checked with data control samples. In particular:
\begin{enumerate}
\item The trigger efficiency evaluated by MC is $99.9$\%, with
excellent  data-MC agreement for the trigger sectors multiplicities.
\item The effects of EVCL and FILFO are  evaluated using a downscaled
  set of non filtered data with less stringent
  cuts  in order to get a ``minimum bias`` sample. On signal events the efficiency
  of the minimum bias selection is 99.88\%.
  We have found that the EVCL procedure
  introduces a signal loss of \ab1.5\%, as  expected also from MC. The corresponding bias on the
  Dalitz plot parameters has been included in the
  systematic error. No bias is introduced by the FILFO procedure.
\item The tracking and vertexing efficiencies have been estimated from
  the data-MC ratio observed in a sample of
  \fipippimpiz~ events with charged pion
  momenta in the same range as those from the \etappp\ decay
  \cite{CAMILLA_NOTE}. These events can be selected with low
  background requiring the detection of the photons associated to the
  $\piz$ in the EMC and only one track in the DC and thus are suited to
  study on data the single charged track reconstruction
  efficiency, $\epsilon_{\rm trk}$, and the charged vertex
  reconstruction efficiency, $\epsilon_{\rm vtx}$:
  $\epsilon^2_{\rm trk} \epsilon_{\rm vtx,\ data}/\epsilon^2_{\rm trk} \epsilon_{\rm vtx,\ MC} = 0.974~ \pm~ 0.006.$
  This ratio is constant  for all momenta, introducing no bias in the Dalitz plot distribution.
  All variables used in the fit are evaluated
  in the $\eta$ rest frame, which in the
  laboratory has a momentum of \ab363 MeV. Therefore to each momentum bin
  in the rest frame corresponds a wider interval
  in the lab; MC-data discrepancies are further diluted by this effect.
\item A correction to the MC detection efficiency for low energy photons has been
  obtained by comparing the photon energy spectrum of a data subsample  to
  the expected MC spectrum; the average correction factor is
  0.964.
\end{enumerate}
The overall selection efficiency, taking into account all the data-MC
corrections is found to be $\epsilon = ( 33.4~ \pm~ 0.2 )$\%.
The expected background contamination, obtained from MC simulation is 0.3\%.

After background subtraction we remain with $ 1.34 \cdot 10^6$ events.

The Dalitz plot density is shown in  fig.\ref{dalitz_plot}.

\FIGURE{
 \figbc dalitz;9;
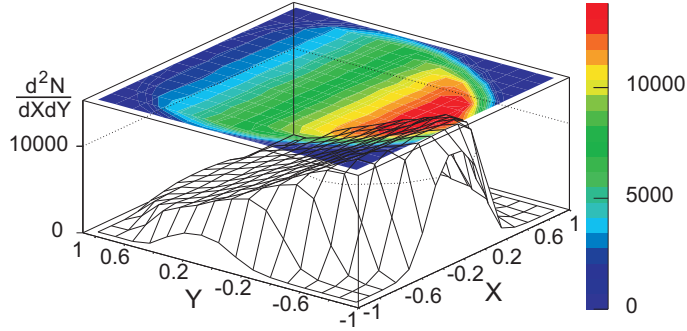
  \caption{ DP distribution for the whole
    data sample. The plot contains $1.34$ millions of events in $256$
    bins.}
  \label{dalitz_plot}}
The signal selection efficiency $\epsilon\left(X,Y\right)$ as function of
the DP point
is obtained by MC, for each $\left(X,Y\right)$
bin, as the ratio:
\begin{equation}
\epsilon\left(X,Y\right) = \frac{N_{\rm rec}\left(X,Y\right)}{N_{\rm gen}\left(X,Y\right)}
\end{equation}
where $N_{\rm rec,\:gen}(X,Y)$ are respectively the reconstructed and generated DP populations.
This approach accounts for resolution effects as long as  the
MC correctly reproduces the Dalitz plot shape;  a first estimate
of the Dalitz plot parameters to be used in the final MC was obtained from
a preliminary fit to a data subsample.
The efficiency $\epsilon\left(X,Y\right)$ has a smooth behavior all over the entire DP.
The projections of  $\epsilon(X,Y)$ are shown in  fig.\ref{effXandY}.
\FIGURE{
\cl{\figb effx;6.3;\kern.5cm\figb effy;6.3;}
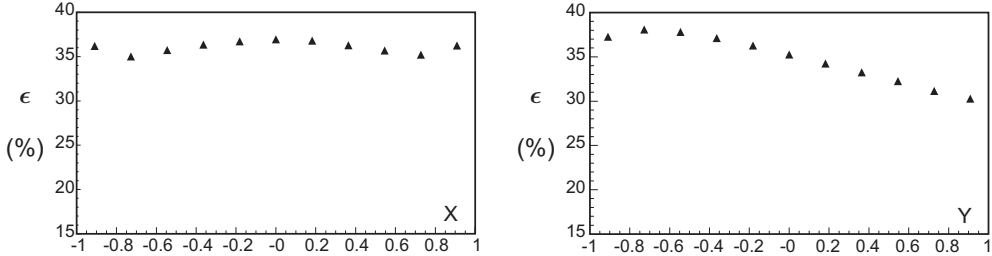
\caption{ Left: Efficiency vs $X$. Right: Efficiency
    vs $Y$.}
\label{effXandY}}
While the efficiency appears to be rather flat on $X$ (and symmetric as expected), it decreases approximately
linearly  with $Y$.
In fact a large Y value means a low-momentum $\pi^{\pm}$
in the decay to which corresponds  a lower tracking/vertexing efficiency.
The resolutions from MC on the DP variables $\left(X,Y\right)$ are  shown in
fig.\ref{resXandY}.
The $Y$ variable, which is proportional to
the $\piz$ kinetic energy, is evaluated \cite{NOTA215} as the average between the
``direct'' determination obtained from the energy and direction of the
two clusters associated to the $\piz\to\gamma\gamma$ decay and the
``indirect'' determination :
$
T_{0} =  M_{\eta} - \left(E_{\pip}+E_{\pim}\right) - M_{\piz}
$.
Due to our excellent momentum resolution for charged tracks, the core of both distributions in fig.\ref{resXandY}
can be fitted with a gaussian with $\sigma = 0.02$.
\FIGURE{
\cl{\figb resX_FIT;6.3;\kern.5cm\figb resY_FIT;6.3;}
\caption{$X$
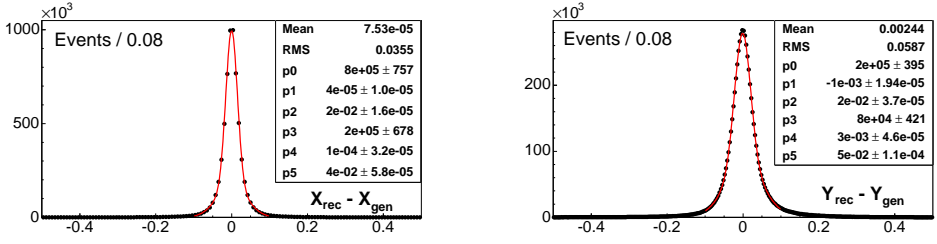
    (left) and $Y$ (right) resolution from MC.}
\label{resXandY}}

\section{Fit of Dalitz plot}
The expected Dalitz density is taken as:
\begin{equation}
\Gamma(X,Y) =|A(X,Y)|^2= N (1 + aY + bY^{2} + cX + dX^{2} + e XY + ...).
\label{eq:standardpar}
\end{equation}
with N being a normalization constant.
The fit to the Dalitz plot is done in two dimensions, minimizing the $\chi^2$ function.
Bins intersecting the Dalitz plot boundary are not included in the fit.
The fit procedure has been tested on MC by verifying that the fit
reproduces
in output the same input values of the DP parameters.

The fit results for different forms of  $|A|^2$   and for  $\Delta X$=$\Delta Y$=0.125
  (154 bins fitted) are shown in  Table \ref{tab:tot_stat}.
A fit with only quadratic terms gives a very low C.L. of  ${\mathcal O}(10^{-6})$ or less.
Including cubic terms as
\begin{equation}\kern-3mm\Gamma(X,Y)=N(1\!+\!aY\!+\!bY^2\!+\!cX\!+\!dX^2\!+\!eXY\!+\!fY^3\!+\!gX^3\!+\!hX^2Y\! +\! lXY^2)
\label{eq:nostandardpar}
\end{equation}
\noindent
results in much better fits with C.L. $>$ 70 \%.
In particular the coefficients $f$  of the $Y^3$ term and $d$ of the $X^2$
term are clearly
required while the other ones ($g, h, l$) turn out to be consistent with zero.
\renewcommand\arraystretch{1.2}
\TABLE{
  \begin{tabular}{|c|c|c|c|c|c|c|c|}
    \hline
    dof & CL & $a\!\x\!10 ^3$ & $b\!\x\!10 ^3$&$c\!\x\!10 ^3$ &$d\!\x\!10 ^3$&$e\!\x\!10 ^3$&$f\!\x\!10 ^3$\\
    \hline
    147 & 73\% & $-$1090$\pm$5 & 124$\pm$6 &
    2$\pm$3  & 57$\pm$6 & $-$6$\pm$7 & 140$\pm$10\\
    \hline
    149 & 74\% & $-$1090$\pm$5 & 124$\pm$6 &
    & 57$\pm$6 & & 140$\pm$10   \\
    \hline
    150 & $<10^{-6}$ & $-$1069$\pm$5 & 104$\pm$5 &
    &  & & 130$\pm$10 \\
    \hline
    150 & $<10^{-8}$ & $-$1041$\pm$3 & 145$\pm$6 &
    & 50$\pm$6  & &  \\
    \hline
    151 & $<10^{-6}$ & $-$1026$\pm$3 & 125$\pm$6 &
    &  & &  \\
    \hline
  \end{tabular}
  \caption{Fits for different forms of $|A|^2$. We take row two as our result.}
  \label{tab:tot_stat}}
As expected from \C-invariance $c$ and $e$ are consistent with zero. Ignoring them in the fit does not affect the other parameters.
Our final results for the Dalitz plot parameters are those shown in second
row of the table. The corresponding correlation coefficients are shown in eq.(\ref{corr}).
Fig \ref{agree_X_Y} and Fig \ref{residual} show respectively a comparison
between fit and data for the projections in $X$ or $Y$ and the normalized residuals as function of bin number (left) and DP variables (right).\\
\FIGURE{
\cl{\figb agreex;6;\kern.5cm\figb agreey;6;}
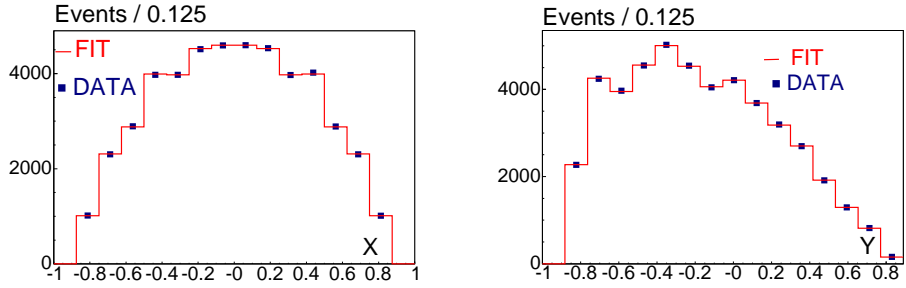
\caption{ Comparison between data(points) and fit(histogram)
  for X,Y projections of the Dalitz plot distribution.}
\label{agree_X_Y}}
\FIGURE{
\cl{\figb Nbin_residuals;6;\kern.5cm\figb DP_residuals;6;}
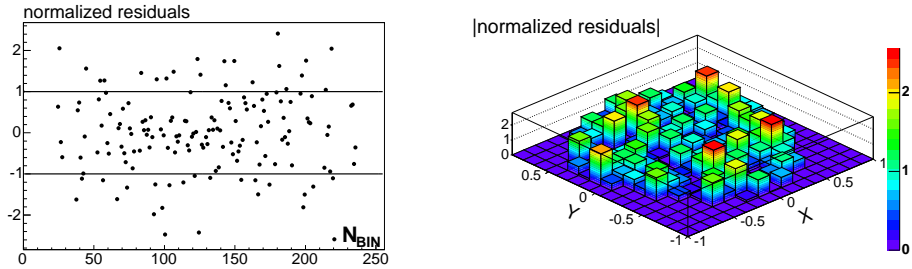
\caption{Left: Distribution of normalized residuals as function of bin number. Residuals fluctuate around zero, with 44 out of 154 exceeding 1, in absolute value. Right: Absolute value of normalized residuals distribution as function of $X$ and $Y$.}  
\label{residual}}
\section{Systematic uncertainties \label{Systematics}}
We have estimated the systematics errors due to the following sources:
\begin{description}
\item[Analysis cuts]
We have moved separately the following cuts: $\theta_{\gamma\gamma}$ in the range
$[15\deg, 21\deg]$ with a step of $3\deg$, $P(\chi^{2})$ in the range
[0.01, 0.15] with a step of 0.05, $E_{\gamma}$ in the range $[10, 25]
\MeV$ with a step of 5 MeV and $\sum E_{\gamma}$ in the range $[780,
820]$ MeV with a step of 10 MeV.
We find a negligible effect on the parameter estimates.
\item[Efficiency]
All reconstruction efficiencies have been checked with
data, using control samples.
We find excellent agreement between data and MC for various kinematical
distributions (see fig.\ref{fig_pip_pim_fot_data_mc}). 
Concerning the photon detection efficiency we have checked that the error with which we estimate the ratio 
$\epsilon_{data}/\epsilon_{MC}$ has a negligible impact on the estimate DP slope parameters.
Only the EVCL procedure gives observable effects, as verified with the minimum bias sample.
\item[Resolution and binning]
Energy resolution for the photons is checked by comparing E$_{\gamma}$  distributions
after the kinematic fit on data and MC. We find good agreement over the entire
distributions. The drift chamber momentum resolution and absolute scale is
checked run by run with the reconstructed $K_S$ mass from
$\K_S\to\pi^+\pi^-$ events. Binning size was changed up to a factor of two: $0.11<\Delta X,\,\Delta Y<0.2$.
\item[Background contamination]
The main source of backgrounds  are:
$\fietag$ with \etappp\ ,$\piz\rightarrow e^{+}e^{-}\gamma$ and
$\phi\rightarrow\omega\piz$ with $\omega\rightarrow\pip\pim\piz$.
Changing the cut on $m_{\gamma\gamma}$ in a wide range,
corresponding to a background change from 0.7\% to 0.2\% ,  we find small changes for the parameter values.
\item[Stability with respect to data taking conditions]
We have divided our data sample in 9 periods of about 50 \pbinv ~each.
We find that the results for each parameter are consistent with no change.
\item[Radiative corrections]
We have generated $10^7$
$\etappp\gamma$ decays, according to ref. \cite{Gatti}. The bin by bin
ratio of the DP density for
$\etappp\gamma$ decays to that for \etappp\ decays
can be fitted with a constant with $\chi^{2}$/dof = $154/153$ corresponding to a CL of
 46\%.
\end{description}
The results are shown in Table \ref{tab:system_tot}.
\TABLE{
    \begin{tabular}{|c|c|c|c|c|}
      \hline
      Source &   $\Delta a$ & $\Delta b$ & $\Delta d$ & $\Delta f$ \\
      \hline       EVCL & $-$0.017 & 0.005 & $-$0.012 & 0.01  \\
      \hline       binning &  $-$0.008 +0.006 & $-$0.006 +0.006 & $-$0.007 +0.001 & $-$0.02 +0.02\\
      \hline       background &  $-$0.001 +0.006 & $-$0.008 +0.006 & $-$0.007 +0.007 & $-$0.01 \\
      \hline       Total  &  $-$0.019 +0.008 & $\pm$ 0.010 & $-$0.016 +0.007 & $\pm$ 0.02 \\
      \hline
    \end{tabular}\\[2pt]
  \caption{ Summary of
      the systematic errors on the Dalitz plot parameters.}
  \label{tab:system_tot}}
\noindent
For each effect mentioned above the systematic error has been
estimated as the maximum parameter variation with respect to
the reference value; the total systematic error is the sum in
quadrature of the different contributions.
\FIGURE{
\figbc 4figs;13;
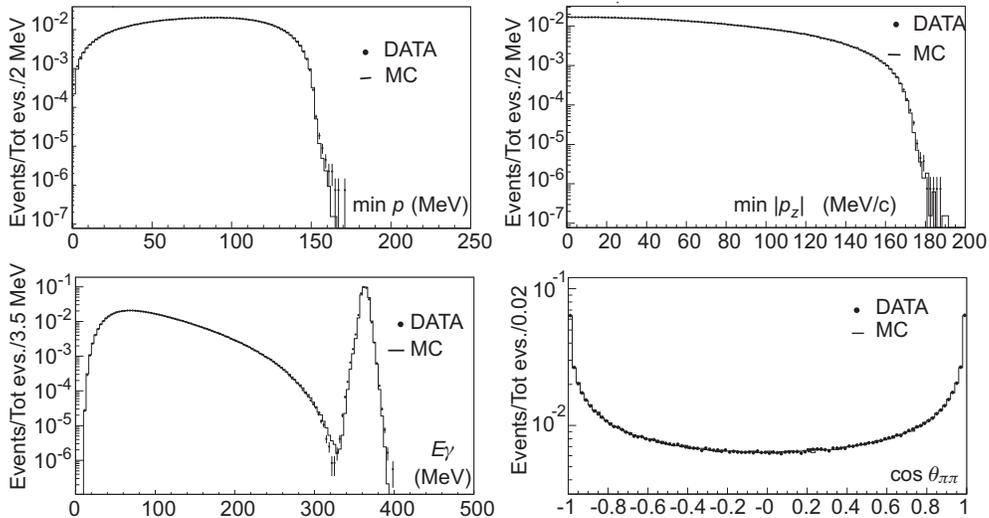
\caption{ Data vs Monte Carlo comparisons in log scale.
Clockwise from top left: minimum $p_T$ and $|p_z|$,
 $\cos{\theta}$ between pion  tracks and $E_{\gamma}$ for photons.}
\label{fig_pip_pim_fot_data_mc}}
\section{An alternative parametrization of the decay amplitude}
We have also fitted the Dalitz plot with a different parametrization
which takes into account the final state $\pi$-$\pi$ rescattering.
Since strong interactions are expected to mix
the two isospin $I $= 1 final states of the \etatrepi~ decay, it is possible to
introduce a unique rescattering matrix $R$ which mixes the corresponding
$I$=1 decay amplitudes  \cite{D'Amb_Isi} , for which we have:
\begin{equation}
\pmatrix{A^{(1)}_{+-0} \cr\noalign{\vglue-5pt}\cr
 A^{(1)}_{000}}_R = T_n\, R\, T_n^{-1}\pmatrix{A^{(1)}_{+-0} \cr\noalign{\vglue-5pt}\cr A^{(1)}_{000}}
\end{equation}

where:
\begin{equation}
R=1+i\pmatrix{\alpha&\beta'\cr\alpha'&\beta}\ \ \textrm{and}\ \  T_{n} = \pmatrix{
1   & -1\cr
3 & 0\cr}.
\end{equation}
According to ref. \cite{D'Amb_Isi}, the rescattering phases depend on the $x$ and $y$ variables\footnote{ We define
$x$ and $y$ as:
$x=(s_1-s_2)/m^2_\pi$ and $y=(s_3-s_0)/m^2_\pi$
with $s_i=s,t,u$ for $i=1,2,3$ and $m^2_\pi=(m^2_{\pi^+} + m^2_{\pi^-} + m^2_{\pi^0})/3. $} as
\begin{equation}\eqalign{\alpha &= \alpha_{0} +  {\mathcal O} \left ( x^{2}, y^{2} \right )\cr
\alpha'&= \alpha'_0\,y + {\mathcal O} ( x^2, y^2 )\cr}\kern.5cm
\eqalign{\beta&= \beta_{0} +  {\mathcal O}(x, y)\cr
\beta' &= \beta'_0(y^2+ x^2/3)/y + {\mathcal O}(x^2, y^2)\cr}
\end{equation}
where  $\alpha_{0}= 0.18$, $\alpha^{'}_{0}= -0.11$, $\beta_{0} = 0.06$,
 $\beta^{'}_{0}=-0.022$ are obtained from \cite{D'Amb} after proper rescaling from kaon to $\eta$ mass.
The complete amplitudes,
keeping the expansion in powers of $x$ and $y$ up to quadratic terms,
are then given by:
\begin{equation}\eqalign{\kern-.8cm
    (A_{+-0})_R&= \bar a  (1 + i\alpha_0)- \left(\bar b(1 + i\beta_0 ) + i\alpha'_0\bar a\right) y +\left(\bar c (1 + i\alpha_0)-\bar d(1 + i\beta_0)\right.\cr
    &\kern1cm  \left. + i\beta'_0\bar b\right) y^2 +
    \left(\bar c(1 + i\alpha_0) +\bar d (1 + i\beta_0) + i\beta'_0\bar b\right) x^2/3\cr}
  \label{+-0}
\end{equation}
and
\begin{equation}
  (A_{000})_R= 3 ~\bar{a} (~1 + i\alpha_{0}~) + [~3 ~\bar{c} (1 + i~\alpha_{0}) + 3i~\beta^{'}_{0} ~\bar{b}~] (x^{2}/3.+ y^{2})
\end{equation}
We have fitted the Dalitz plot with the above parametrization and
the fit results are given in Table \ref{tab:newres}.
\TABLE{
  \begin{tabular}{|c|c|c|c|c|c|}
    \hline
    dof & $P_{\chi^{2}}$ & $\bar a$\x1000 &  $\bar b$\x1000 & $\bar c$\x1000 & $\bar d$\x1000 \\
    \hline
    150 & $56\%$ &$-$71.12\plm 0.07$^{+0.08}_{-0.23}$ & 13.71\plm 0.04$^{+0.06}_{-0.27}$ & 0.46\plm0.03$^{+0.13}_{-0.08}$ & $-$0.76\plm0.02$^{+0.02}_{-0.04}$\\
    \hline
  \end{tabular}\\
  \caption{ Results of the fit with a parametrization of the form eq.~\ref{+-0}.    }
  \label{tab:newres}}
The systematic uncertainty on the parameters has been evaluated as described in section \ref{Systematics}.\\
From the above results it is possible to extract the Dalitz plot
slope $\alpha$ of the
\etapizpizpiz~ decay. From its definition:
 $$|A_{000}|^2\propto1 + 2 \alpha z  ~~~~~~~~; ~~~~~~~z =9\, m^4_\pi/(4\,
 m^2_\eta\,Q^2)\x(x^2/3+ y^2)$$
 we get:
\begin{equation}
 \alpha= \frac{4~ m^{2}_{\eta}~Q^{2}}{9~ m^{4}_{\pi}}\frac{[~
 \bar{c}~(1 + \alpha_{0}^{2})
 +~\beta^{'}_{0} ~\alpha_{0} ~\bar{b} ~]} {~\bar{a}~ (~1 + \alpha_{0}^{2}~)} =
- 0.038 ~\pm 0.003 (\rm stat)^{ + 0.012} _{-0.008}  (syst)
\end{equation}
in agreement with the PDG \cite{PDG} average $\alpha=-0.031\pm 0.004 $
and the recent KLOE preliminary result
$\alpha=-0.027\pm 0.004^{ + 0.004} _{-0.006}$  \cite{LEPTON07}.\\
\section{Asymmetries}
While the polynomial fit of the Dalitz plot density gives valuable
information on the matrix element,   integrated asymmetries are
very sensitive in assessing the possible presence of \C\ violation
in amplitudes of given $\Delta I$.
In particular left-right asymmetry -
related to the $c$ parameter in our fit - tests \C\ violation with no
specific $\Delta I$ constraint;
quadrants asymmetry tests \C\ violation for $\Delta I = 2$ and sextants
asymmetry (for a  definition see ref. \cite{Layter})
 tests \C\ violation for $\Delta I = 1$.

For this measurement care must be taken of possible slight differences
 between $\pi^+$ and $\pi^-$ reconstruction efficiencies. To this aim
we estimate the MC efficiency separately for each region of the Dalitz plot,
as the ratio between reconstructed and generated events in the region.
This definition takes into account the resolution effects as well.
From a sample of $5.7 \times 10^{6}$ MC events we get:

\begin{tabular}{lcc}
$\epsilon_L    \!= \!(34.91\!\pm\!0.02)\%$ & $\epsilon_R     =(35.05 \pm 0.02)\% $\\
$\epsilon_{13} \!= \!(35.01\!\pm\!0.02)\%$ & $\epsilon_{24}  =(34.95 \pm 0.02)\% $ & (quad.)\\
$\epsilon_{135}\!= \!(35.00\!\pm\!0.02)\%$ & $\epsilon_{246} =(34.96 \pm 0.02)\% $ & (sext.)\\
\end{tabular}
\def\xn{\!\times\!}\\
We have checked these values estimating
 the asymmetries on Monte Carlo: these
turn out to be all compatible with zero. We then evaluate the
asymmetries on data by  subtracting  the MC expected background and
correcting the  ``raw'' asymmetries with the above efficiencies. We
obtain:
$$ A_{LR}= (9 \pm 10 )\xn 10^{-4},\kern5mm
A_{Q}  = (-5 \pm 10)\xn 10^{-4},\kern5mm
A_{S}  = (8 \pm 10)\xn 10^{-4}.$$
Systematic uncertainties on the asymmetries are obtained from studying:
 a) sensitivity to background, by varying cuts,
 b) event selection (EVCL) by use of the minimum bias sample and
 c) MC-data comparison using $\phi\to\pi^+\pi^-\pi^0$ events.
In particular the tracking efficiency has been evaluated separately
for the two charges, since in the MC a small
but statistically significant difference in left and right
efficiencies is evident.
The above difference is due to a slightly different tracking efficiency vs
 $p_T$ for positive and negative pions because of nuclear interactions.

Since we require both tracks to be reconstructed
the absolute value of the efficiency is not important for the
asymmetry, but rather its dependence upon the pion momentum. The
good data-MC agreement has been already demonstrated for both charges
on the signal. We here use the    $\phi \to \pi^+\pi^-\pi^0$
 control sample to check the agreement between data and MC for the
$\pi^+$ and $\pi^-$ efficiencies as a function of momentum (see
fig.\ref{Eff_charges}).

The control sample agrees well
with MC within errors, and the data-MC ratio is well fitted by a
constant.
\FIGURE{
\figbc efftrk;8;
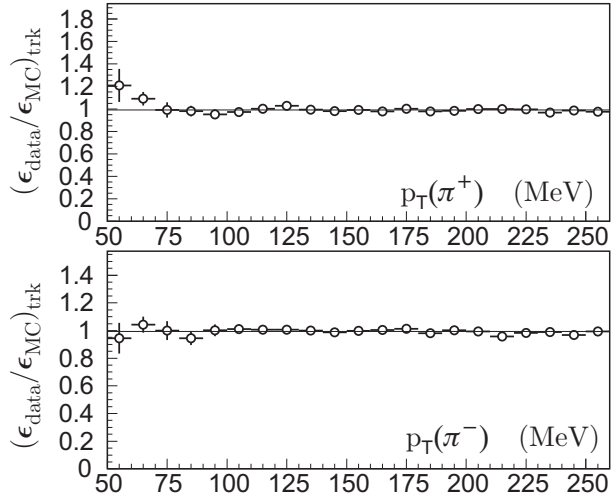
\caption{ The data-MC ratio of tracking efficiency for
    $\pi^+$ (top) and $\pi^-$ (bottom) vs pion $p_T$. }
\label{Eff_charges}}

In order to assess the possible systematic uncertainties connected with the tracking
efficiencies we adopt a conservative approach: we estimat
the maximum positive or negative linear slopes compatible within one sigma with the
fit of the distributions shown in fig.\ref{Eff_charges}. Then we
have assumed that the two charges behave with  {\em opposite}
slopes. This gives us two possibilities: $\pi^+$ with positive slope
and $\pi^-$ with negative slope or  vice-versa. We have then
reweighted the events according to these two possibilities and used
the maximum difference observed in the asymmetries as the corresponding
systematic error.
The systematics connected with the
asymmetries are shown in Table \ref{tab:asym_tot}.
\TABLE{
\vspace{+20 pt}
     \begin{tabular}{|l|c|c|c|}
       \hline
       Syst. Effect &Left-Right &    Quadrant &  Sextant \\
       \hline
       Background & $(-0.2 / +0.1)\times 10^{-3}$ & $(-0.2/ +0.2)\times 10^{-3}$ &$( +0.3)\times 10^{-3}$\\
       EVCL & $(-0.5 )\times 10^{-3}$ & $(-0.3)\times 10^{-3}$ &$( +0.7)\times 10^{-3}$\\
       Efficiency&  $(-1.3 / +0.9)\times 10^{-3}$ & $(-0.3/ +0.2)\times 10^{-3}$ &$( -1.3)\times 10^{-3}$\\
       Total & $(-1.4 / +0.9)\times 10^{-3}$ & $(-0.5/ +0.3)\times 10^{-3}$ &$( -1.3/+0.8)\times 10^{-3}$\\
       \hline
     \end{tabular}\\
   \caption{Systematic errors on asymmetries. }
   \label{tab:asym_tot}}
Therefore the final results for the asymmetries are:
$$\eqalign{
A_{\rm LR}& = (+0.09 \pm 0.10~ ^{+0.09}_{-0.14}  )\times 10^{-2}\cr
A_{\rm Q}& =  (-0.05 \pm 0.10~ ^{+0.03}_{-0.05}  )\times 10^{-2}\cr
A_{\rm S}&=   (+0.08 \pm 0.10~ ^{+0.08}_{-0.13}  )\times 10^{-2}.\cr}$$
where the first (second) is the statistical (systematic) error.

\section{Conclusions}
The results including the statistical uncertainties coming from the
fit and the estimate of systematics are:
\begin{equation}\eqalign{
a &= -1.090 \pm 0.005 (\rm stat) ^{+ 0.008}_{- 0.019} (syst)\cr
b &= 0.124 \pm 0.006 (\rm stat) \pm 0.010 (syst)\cr
d& = 0.057 \pm 0.006 (\rm stat) ^{+ 0.007}_{- 0.016} (syst)\cr
f &= 0.14 \pm 0.01 (\rm stat) \pm 0.02 (syst)}
\end{equation}
Below we give the normalized correlation coefficients for the DP parameters.
\begin{equation}\matrix{
     & a      & b & d & f \cr
a   & 1     & -0.226 & -0.405 & -0.795\cr
b   &        & 1         &  0.358 &  0.261 \cr
d    &       &            &  1        &  0.113 \cr
f    &       &             &            & 1      \cr}
\label{corr}
\end{equation}
\noindent
The following comments are in order:
\begin{enumerate}

\item the fitted value for the quadratic slope in $Y$ is almost one half of
  the simple current algebra prediction ($b = a^{2}/4$), thus calling for
  significant higher order corrections;

\item the quadratic term in $X$ is unambiguously different from
  zero;

\item similarly for the large cubic term in $Y$;

\item the fit results show correlations between the DP
  parameters. This should be properly taken into account for a correct
  error estimate when integrating the amplitude over phase space to get the decay
 width;

\item fitting the $\eta \to \pip \pim \piz $ Dalitz plot with an alternative parametrisation we
  obtained a prediction for the $\eta \to 3 \piz $ slope which is
  consistent with the PDG average and the KLOE measurement;

\item we do not observe any evidence for \C\ violation in
the $\eta\to\pip\pim\piz$ decay since the $c$ and $e$ parameters of the
Dalitz plot and the charge
asymmetries are all perfectly consistent with zero.

\end{enumerate}

\section*{Acknowledgements}
We thank G. D'Ambrosio for many fruitful discussions.
We thank the DAFNE team for their efforts in maintaining low background running
conditions and their collaboration during all data-taking.
We want to thank our technical staff:
G.F. Fortugno and F. Sborzacchi for their dedicated work to ensure an
efficient operation of
the KLOE Computing Center;
M. Anelli for his continuous support to the gas system and the safety of
the
detector;
A. Balla, M. Gatta, G. Corradi and G. Papalino for the maintenance of the
electronics;
M. Santoni, G. Paoluzzi and R. Rosellini for the general support to the
detector;
C. Piscitelli for his help during major maintenance periods.
This work was supported in part
by EURODAPHNE, contract FMRX-CT98-0169;
by the German Federal Ministry of Education and Research (BMBF) contract 06-KA-957;
by the German Research Foundation (DFG),
'Emmy Noether Programme', contracts DE839/1-4;
by INTAS, contracts 96-624, 99-37;
and by the EU Integrated Infrastructure
Initiative HadronPhysics Project under contract number
RII3-CT-2004-506078.

\catcode`@=11
\catcode`\%=12
\catcode`\|=14
\newcommand\epja[3]  {\@spires{EPHJA
		{{\it Eur.\ Phys.\ J. }{\bf A #1} (#2) #3}}
\renewcommand\jphg[3]   {\@spires{JPHGB
        {{\it J. Phys.\ }{\bf G #1} (#2) #3}}
\catcode`\%=14
\catcode`\|=12
\catcode`@=12

\bibliographystyle{JHEP}

%

\begin{thebibliography}{99}

\bibitem{Sutherland}
  J.S. Bell and D.G. Sutherland, Current algebra and $\eta\to3\pi$, \npb{4}{1968}{315}

\bibitem{BijGa02}For a review see
  J. Bijnens and J. Gasser, Eta Decays and Beyond $p^{4}$ in Chiral Perturbation Theory, Physica Scripta  T99, (2002) 34.

\bibitem{PDG}
  Particle Data Group Collaboration, W.~M.~Yao {\it et al.},  \jphg{33}{2006}{1}.

\bibitem{GasLut85}
  J.~Gasser and H.~Leutwyler, Chiral perturbation Theory to one loop, Annals Phys. {\bf 158} (1984) 142;
  Eta to 3 $\pi$ to one loop,  \npb {250} {1985} {539}.

\bibitem{BijGho07}
  J.~Bijnens and K.~Ghorbani, $\etatrepi$ at Two Loops in Chiral Perturbation Theory, \jhep{0711} {2007} {030}.

\bibitem{Borasoy}
  B.~Borasoy and R.~Nissler, Hadronic $\eta$ and $\etap$ decays, \epja{26}{2005}{383}.

\bibitem{DAFNE}
  DAFNE Collaboration, M.~Zobov  {\it et al.}, DAFNE status and upgrade plans, arXiv:
  \arXivid{0709.3696}.

\bibitem{DC}
  KLOE Collaboration, M.~Adinolfi {\it et al.}, The tracking detector of the KLOE experiment, \nim{A488}{2002}{51}.

\bibitem{EMC}
  KLOE Collaboration, M.~Adinolfi {\it et al.}, The KLOE electromagnetic calorimeter, \nim{A 482} {2002}{364}.

\bibitem{TRIGGER}
  KLOE Collaboration,  M.~Adinolfi {\it et al.}, The trigger system of the KLOE experiment,
  \nim{A 492} {2002}{134}.

\bibitem{EVCL}
  KLOE Collaboration,  F.~Ambrosino {\it et al.}, Data handling, reconstruction, and simulation for the KLOE experiment, \nim{A 534} {2004}{403}.

\bibitem{CAMILLA_NOTE}
  C.~Di Donato, KLOE Note 214, \href{http://www.lnf.infn.it/kloe/pub/knote/kn214.ps}{$\Phi\rightarrow\eta' \gamma$ in $\pi^+ \pi^- 7 \gamma$ at KLOE}.

\bibitem{NOTA215}
  F.~Ambrosino, T.~Capussela, F.~Perfetto, KLOE Note 215, \href{http://www.lnf.infn.it/kloe/pub/knote/kn215.ps}{Dynamics of   $\etappp$}.

\bibitem{Gatti}
  C.~Gatti, Monte Carlo simulation for radiative kaon decays, \epjc{45} {2005} {417}.

\bibitem{D'Amb_Isi}
  G.~D'Ambrosio, G.~Isidori, CP violation in kaon decays, \ijmpa{13}{1998}{1}.

\bibitem{D'Amb}
  G~D'Ambrosio, G.~Isidori, A.~Pugliese and N.~Paver, Strong rescattering in decays $K\rightarrow3\pi$ and low-energy meson interactions, \prd {50}{1994}{5767}.

\bibitem{LEPTON07}
  KLOE Collaboration,  F.~Ambrosino {\it et al.}, Measurement of the slope parameter $\alpha$ for the $\eta\rightarrow3\pi^0$ decay at KLOE, \arXivid{0707.4137}.

\bibitem{Layter}
  J. G. Layter {\it et al.}, Measurement of the charge Asymmetry in the decay $\etappp$, \prl{29} {1972}{316}.


\end{thebibliography}
%
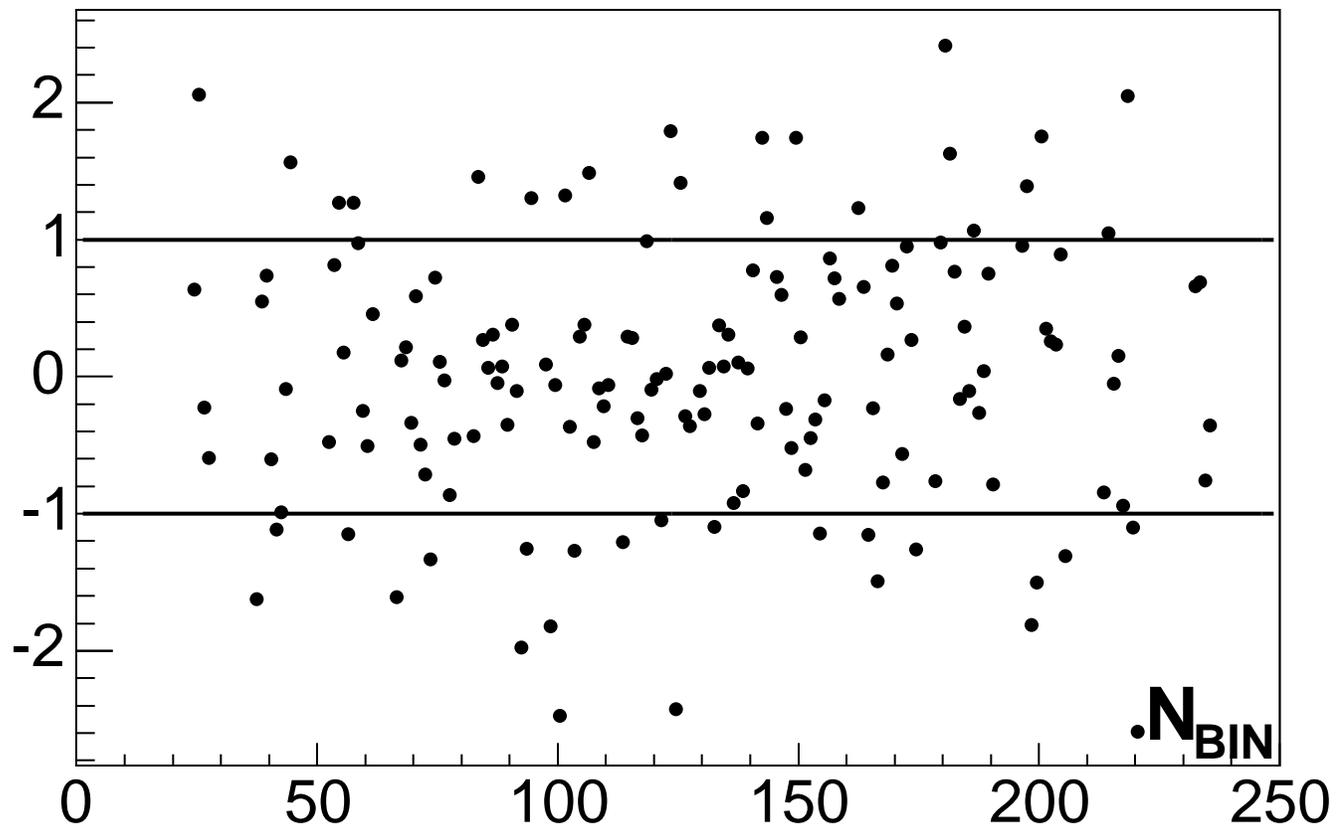
\end{document}